%% file: main.tex
\documentclass[a4paper]{article}

\input{definitions}

\input{00.decl.tex}

\title{A Systematic Comparison of Phonetic Aware \\ Techniques for Speech Enhancement}
\name{Or Tal$^1$, Moshe Mandel$^1$, Felix Kreuk$^{2}$, Yossi Adi$^{1,2}$}
\address{
  $^1$Hebrew University of Jerusalem\\
  $^2$Meta AI Research}
\email{or.tal1@mail.huji.ac.il}

\begin{document}

\input{01.abstract.tex}
\input{01.intro.tex}
\input{03.model.tex}
\input{04.exp.table.tex}
\input{04.exp.tex}
\input{04.exp.x-factor.table.tex}
\input{02.related.tex}
\input{05.conclusion.tex}

\bibliographystyle{IEEEtran}
\bibliography{mybib}
\end{document}

%% file: definitions.tex
\usepackage{bm}
\usepackage{amssymb,amsmath,graphicx,bbm,color,booktabs}
\usepackage{amsopn}

\newcommand{\vx}{\bm{x}}               
\newcommand{\vy}{\bm{y}}       \newcommand{\vyh}{\hat{\bm{y}}}

\newcommand{\veta}    {\bm{\eta}}

\newcommand{\R}{\mathbb{R}}

\renewcommand{\eqref}[1]{Eq.~(\ref{#1})}

%% file: 00.decl.tex
\usepackage{INTERSPEECH2022}
\usepackage{dsfont}
\usepackage{color}
\usepackage{graphicx}
\usepackage{multicol}
\usepackage{multirow,makecell, rotating}
\usepackage[table,xcdraw]{xcolor}
\usepackage{makecell}
\usepackage{acronym}
\usepackage{pifont}
\graphicspath{ {./LaTeX/figures }}
\usepackage[breaklinks]{hyperref}

\acrodef{SNR}{Signal-to-Noise Ratio}
\acrodef{ASR}{Automatic Speech Recognition}
\acrodef{SSL}{Self-Supervised Learning}

%% file: 01.abstract.tex
\maketitle
\begin{abstract}
Speech enhancement has seen great improvement in recent years using end-to-end neural networks. However, most models are agnostic to the spoken phonetic content. Recently, several studies suggested phonetic-aware speech enhancement, mostly using perceptual supervision. Yet, injecting phonetic features during model optimization can take additional forms (e.g., model conditioning). In this paper, we conduct a systematic comparison between different methods of incorporating phonetic information in a speech enhancement model. 
By conducting a series of controlled experiments, we observe the influence of different phonetic content models as well as various feature-injection techniques on enhancement performance, considering both causal and non-causal models. 
Specifically, we evaluate three settings for injecting phonetic information, namely: i) \emph{feature conditioning}; ii) \emph{perceptual supervision}; and iii) \emph{regularization}. Phonetic features are obtained using an intermediate layer of either a supervised pre-trained \ac{ASR} model or by using a pre-trained \ac{SSL} model. We further observe the effect of choosing different embedding layers on performance, considering both manual and learned configurations. 
Results suggest that using a \ac{SSL} model as phonetic features outperforms the \ac{ASR} one in most cases. Interestingly, the conditioning setting performs best among the evaluated configurations.
Code is available on the following ~{\color{purple}\href{https://github.com/slp-rl/SC-PhASE}{repository}}.
\end{abstract}
\vspace{-0.2cm}

%% file: 01.intro.tex
\section{Introduction}
\label{sec:intro}
Speech enhancement is the task of maximizing the perceptual quality of speech signals, in particular by removing background noise. Most recorded conversational speech signals contain some form of noise that hinders intelligibility, such as outdoor noises. Thus, speech enhancement is a particularly important task for audio and video calls \cite{reddy2019scalable}, hearing aids \cite{reddy2017individualized}, and can also help automatic speech recognition (ASR) systems \cite{zorila2019investigation}. 

Previous work in speech enhancement showed feasible solutions which estimated the noise model and used it to recover noise-deducted speech~\cite{lim1979enhancement,ephraim1984speech}. Although those approaches can generalize well across domains, they still have trouble dealing with common noises such as non-stationary noise or a babble noise which is encountered when a crowd of people are simultaneously talking. The presence of such noise types degrades hearing intelligibility of human speech greatly~\cite{krishnamurthy2009babble}. 
Recently, deep neural networks (DNN) based models perform significantly better on such noise types while generating higher quality speech in objective and subjective evaluations over traditional
methods~\cite{pascual2017segan,phan2020improving, defossez2020real,fu2021metricgan+, tzinis2021continual}. 
Additionally, deep learning based methods have also shown to be superior over traditional methods for the related task of a single-channel source separation~\cite{chazan2021single,demucs}.

Traditionally, such methods are agnostic to the spoken content in the input speech signals. Recently, several studies suggested incorporating phonetic information into speech enhancement models, namely phonetic-aware speech enhancement. The authors of~\cite{hsieh2020improving} 
suggested using a feature matching loss over representations obtained by self-supervised methods as perceptually-guided loss functions, while the authors of~\cite{von2020multi} 
suggested using an automatic speech recognition based system for a similar purpose. Lastly, the authors of~\cite{du2020pan} proposed to inject supervised phonetic information either as an additional loss function or as another conditioning to the enhancement model. 

In this work, we perform a systematic comparison in order to better comprehend the effect of injecting phonetic information and the mechanisms for effectively incorporating it. We perform a controlled study in which we evaluate both \ac{SSL} and \ac{ASR} as phonetic representations, considering both causal and non-causal setups of speech enhancement models. Specifically, we evaluate three methods for injecting phonetic information during training: i) \emph{supervision}; ii) \emph{regularization}; and iii) \emph{conditioning} (for the non-causal setting only). 

Under the \ac{SSL} setting, previous studies~\cite{lakhotia2021generative, hsu2021hubert, chang2021distilhubert}
considered using different intermediate layers to better capture phonetic information. To mitigate that, we additionally explore the effect of layer selection on model performance, considering both manual selection and learned configurations. 
Results suggest that overall the \ac{SSL} model was more effective than the \ac{ASR} alternative. 
Interestingly, there is notable improvement under the non-causal setup yet only marginal changes under the causal setup. Finally, the conditioning setting was found to be superior to both regularization and supervision in most cases. 

\begin{figure*}[t!]
\centering
\includegraphics[width=0.75\textwidth]{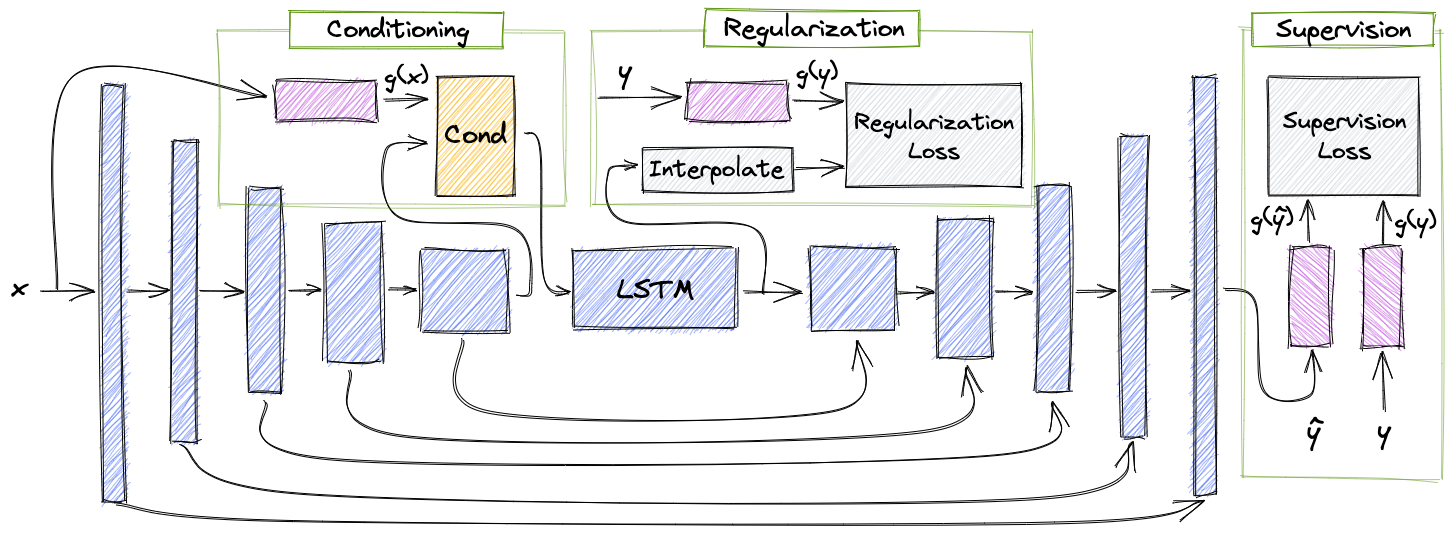}
\vspace{-1em}
\caption{An illustration of the three purposed phonetic settings, the base architecture corresponds to a high level description of Demucs architecture. Each setting was experimented disjointly. Notation: Blue - the baseline model layers, Purple - lexical feature model (frozen through training and inference), Orange - additional linear module for conditioning purpose.}
\label{fig:model}
\vspace{-1.5em}
\end{figure*}

%% file: 03.model.tex
\vspace{-0.2cm}
\section{Model}
\label{sec:model}

\subsection{Problem Setting}
In the task of single channel speech enhancement we are provided with a noisy speech utterance, denoted as $\vx=(x_1,\ldots,x_T)$, where each $x_t\in\R$, $(1\leq t \leq T)$. The length of the speech utterance, $T$, is not a fixed value, as input utterances can have different durations. We assume $\vx$ is comprised of a clean reference signal, $\vy$, and an additive non-stationary noise signal, $\veta$, mixed with an unknown \ac{SNR} such that $\vx = \vy + \veta$. 
We follow the supervised setting, where each noisy input signal, $\vx$, is associated with its clean version $\vy$. Our goal is to learn a function $f$ which maps a given noisy speech to an estimated version of the clean speech. Ideally $f$ should output a prediction vector $\vyh$ which is as close as possible to the reference sample vector $\vy$, i.e. minimizing the distance between $\vyh$ and $\vy$ with respect to a given loss function $\ell$.

\subsection{Model Architecture}
We use the Demucs model architecture~\cite{defossez2019music} as $f$, working directly over the time-domain signal. This model was found to be efficient both for speech enhancement~\cite{defossez2020real} and music source separation~\cite{defossez2019music}. The model consists of a multi-layer convolutional network with a U-Net-like architecture built from three sequential main modules: i) an encoder; ii) a sequential modeling module; and iii) a decoder with corresponding skip connections between the encoder and decoder layers. The noisy audio waveform input is upscaled by an upscaling factor $U$ before being passed into the encoder using a simple sinc interpolation, and downsampled by the same factor after the decoder. The Demucs-based architectures could be then characterized by the upscaling factor $U$, the initial number of hidden channels $H$, and the stride taken in downsampling and upsampling convolutional layers $S$. We distinguish between causal and non-causal setups using a uni-directional or bi-directional LSTMs as sequential modeling accordingly.

\subsection{Phonetic-Aware Settings}\label{phonetic settings}
Consider a phonetic feature vector as an embedded representation of a speech signal which captures its content and intents. 
One may then conclude that using some of the information held within this representation could benefit speech enhancement as it shares prior information regarding the content of the signal. In this work, we explore three modeling settings to inject phonetic information into a speech enhancement model $f$, namely \emph{Regularization}, \emph{Supervision}, and \emph{Conditioning}. See Figure~\ref{fig:model} for a visual description of all settings.

\vspace{0.1cm}
\noindent{\bf Regularization:} 
Formally, let $g$ be a phonetic function which maps a given input to a sequence of $d$-dimensional feature vectors, corresponding to a predefined temporal resolution, e.g one vector per 20ms.
Under the regularization setup we minimize:
\vspace{-0.2cm}
\begin{equation}
	\ell(\vy, f(\vx)) + \lambda \cdot \ell_{reg}(I(e_i(\vx)), g(\vy))
\vspace{-0.2cm}
\end{equation}
$\ell$ is the main training loss, $\ell_{reg}$ is a regularization loss function with a small $\lambda$ coefficient balancing between the two terms in the objective. 
$e_i(\vx)$ is the output of the $i$'th layer of $f$ with respect to the input noisy sample $\vx$, and $I$ is a simple linear interpolation used to match the dimensions of $e_i(\vx)$ and $g(\vy)$. $i$ and $\lambda$ are hyper-parameters. The motivation for this approach is to encourage latent representation to capture phonetic content while removing background noise.

\vspace{0.1cm}
\noindent{\bf Supervision:} 
Under the supervision setting, we use the phonetic model as another perceptual supervision. We augment the training objective with an additional loss term as follows: 
\vspace{-0.2cm}
\begin{equation}
	\ell(\vy, f(\vx)) + 	\lambda \cdot \ell_{sup}(g(f(\vx)), g(\vy))
\vspace{-0.5cm}
\end{equation}
where $\ell_{sup}$ is the supervision loss function, $\lambda$ is a hyper-parameter balancing between the two loss terms in the objective. Notice that unlike the regularization setting in which we minimize the loss function between an intermediate representation from $f$ to the output of $g$, under the supervision setting we back-propagate the gradients through $g$ while optimizing $f$.

\vspace{0.1cm}
\noindent{\bf Conditioning:}
Unlike the two previous settings, under the conditioning setting we do not augment the training objective with additional loss terms. 
Instead, we inject the output of $g$ into an intermediate layer of $f$. Formally, the training objective is $\ell(\vy, f(\vx, g(\vx)))$. 
Practically, we concatenate the representation obtained by the phonetic model with an intermediate representation of the main enhancement model and pass it through by a linear projection layer to reach the expected input dimension. 

\vspace{-0.2cm}
\subsection{Phonetic models}
Roughly two types of phonetic models can be considered, a supervised one in the form of \ac{ASR} and a self-supervised model (e.g., HuBERT~\cite{hsu2021hubert}, WavLM~\cite{chen2021wavlm}, etc.). Both options were shown to be efficient for transcription tasks and thus assumed to capture phonetic information. To evaluate the impact of each model type on the downstream speech enhancement task, we perform a systematic comparison between the two types of models. For the self-supervised setting we use a pre-trained HuBERT model~\cite{hsu2021hubert}, which achieves state-of-the-art results on a large number of tasks~\cite{lakhotia2021generative, yang2021superb}. For the supervised setting we evaluate a pre-trained \ac{ASR} model using a joint CTC attention training objective, similarly to the one proposed in~\cite{watanabe2017hybrid}. Both models were trained on the Librispeech-960 corpus~\cite{panayotov2015librispeech}. Unlike other settings, in the conditioning setting we inject phonetic features during inference. As both phonetic models are non-causal, they will only be applied in the non-causal setup under the conditioning setting.

%% file: 04.exp.table.tex
\begin{table*}[t!]
\centering
\caption{Results for three Demucs configurations using either HuBERT layer-6 (H-L6) or ASR models' output as phonetic features. We report results for a model without phonetic information (Base), Regularization (Reg), Supervision (Sup), and Conditioning (Cond), considering both causal and non-causal setups. }
\vspace{-0.5em}
\resizebox{0.9\textwidth}{!}{%
\begin{tabular}{@{\hskip6pt}l|c|c|ccccc|ccccc} 
    \toprule
    \multicolumn{3}{c}{} & \multicolumn{5}{c}{\bf Causal} & \multicolumn{5}{c}{\bf Non-causal}\\
    \midrule
    & \bf Setting & \bf Phonetic Model & \bf CBAK & \bf COVL & \bf CSIG & \bf PESQ & \bf VISQOL & \bf CBAK & \bf COVL & \bf CSIG & \bf PESQ & \bf VISQOL \\
    \midrule
    \multirow{7}{*}{\rotatebox[origin=c]{90}{H=48,U=4,S=4}}
    & Base & - & 3.47 & 3.67 & 4.34 & 2.95 & 3.20 & 3.50 & 3.65 & 4.33 & 2.93 & 3.20  \\
    & Reg & H-L6 & 3.45 & 3.65 & 4.32 & 2.94 & 3.17 & 3.53 & 3.67 & 4.34 & 2.96 &  3.16 \\
    & Sup & H-L6 & 3.47 & 3.69 & 4.35 & 2.96 & 3.05 & 3.51 & 3.66 & 4.34 & 2.93 & 3.20 \\
    & Cond & H-L6 & - & - & - & - & - & \bf{3.53} & \bf{3.71} & \bf{4.38} & \bf{2.99} & \bf{3.25} \\
    & Reg & ASR & 3.46 & 3.66 & 4.33 & 2.93 & 3.20 & 3.52 & 3.67 & 4.35 & 2.95 & 3.20 \\
    & Sup & ASR & 3.43 & 3.62 & 4.30 & 2.90 & 3.20 & 3.53 & 3.70 & 4.36 & 2.98 & 3.17 \\
    & Cond & ASR & - & - & - & - & - & 3.50 & 3.64 & 4.33 & 2.92 & 3.17 \\
    \midrule
    \multirow{7}{*}{\rotatebox[origin=c]{90}{H=64,U=4,S=4}}
    & Base & - & 3.47 & 3.65 & 4.33 & 2.93 & 3.20 & 3.52 & 3.66 & 4.34 & 2.95 & 3.13 \\
    & Reg & H-L6 & 3.44 & 3.66 & 4.33 & 2.93 & 3.18 & 3.52 & 3.67 & 4.35 & 2.95 & 3.17 \\
    & Sup & H-L6 & 3.45 & 3.65 & 4.33 & 2.93 & 3.19 & \bf{3.54} & \bf{3.72} & \bf{4.38} & \bf{3.00} & 3.21 \\
    & Cond & H-L6 & - & - & - & - & - & 3.52 & 3.69 & 4.35 & 2.98 & \bf{3.27} \\
    & Reg & ASR & 3.45 & 3.64 & 4.32 & 2.92 & 3.19 & 3.53 & 3.68 & 4.35 & 2.97 & 3.18 \\
    & Sup & ASR & 3.47 & 3.67 & 4.34 & \bf{2.95} & 3.21 & 3.50 & 3.65 & 4.33 & 2.93 & 3.18 \\
    & Cond & ASR & - & - & - & - & - &  3.50 & 3.64 & 4.33 & 2.92 & 3.17 \\
    \midrule
    \multirow{7}{*}{\rotatebox[origin=c]{90}{H=64,U=2,S=2}}
    & Base & - & 3.37 & 3.50 & 4.20 & 2.77 & 3.17 & 3.54 & 3.73 & 4.38 & 3.03 & 3.26\\
    & Reg & H-L6 & 3.41 & 3.60 & 4.27 & 2.89 & 3.18 & 3.59 & 3.74 & 4.37 & 3.06 & 3.26\\
    & Sup & H-L6 & \bf{3.43} & \bf{3.62} & \bf{4.29} & \bf{2.91} & 3.19 & 3.56 & 3.72 & 4.36 & 3.04 & 3.24\\
    & Cond & H-L6 & - & - & - & - & - &  \bf{3.62} & \bf{3.81} & \bf{4.44} & \bf{3.12} & \bf{3.32}\\
    & Reg & ASR & 3.45 & 3.62 & 4.30 & 2.90 & 3.20 & 3.58 & 3.75 & 4.41 & 3.06 & 3.32 \\
    & Sup & ASR & 3.43 & 3.59 & 4.26 & 2.89 & 3.21 & 3.56 & 3.72 & 4.36 & 3.03 & 3.28 \\
    & Cond & ASR & - & - & - & - & - & 3.53 & 3.68 & 4.34 & 2.99 & 3.21 \\
    \bottomrule
    \end{tabular}}
\label{tab:main}
\vspace{-1em}
\end{table*}

%% file: 04.exp.tex
\vspace{-0.2cm}
\section{Experiments}
\label{sec:exp}

\subsection{Experimental Setup}
\label{subsec:setup overview}

\noindent {\bf Training Objective.} 
Similar to~\cite{yamamoto2019probability}, we set $\ell$ to be the L1 loss over the waveform together with a multi-resolution STFT loss over the spectrogram magnitudes. Formally, $\ell(\vy,\vyh) = \frac{1}{T}[\|\vy-\vyh\|_1 + \sum_{i=1}^{k}L_{stft}^{(i)}(\vy,\vyh)]$, 
where, 
\begin{equation}
\label{eq:stft}
\begin{split}
&L_{stft}(\vy,\vyh) = L_{sc}(\vy,\vyh) + L_{mag}(\vy,\vyh), \\
&L_{sc}(\vy,\vyh) = \frac{{\||STFT(\vy)| - |STFT(\vyh)|\|}_F}{{\||STFT(\vy)|\|}_F}, \\
&L_{mag}(\vy,\vyh) = \frac{{\|\log|STFT(\vy)| - \log|STFT(\vyh)|\|}_1}{T}.
\end{split}
\end{equation}
in which $\|\cdot\|_1, \|\cdot\|_F$ are the L1 and Frobenius norms. We apply $k$ STFT losses for which $L_{stft}^{(i)}(\vy,\vyh)$ corresponds to a single setting among the union of settings defined by the number of FFT bins $\in\{512, 1024, 2048\}$, hop size $\in\{50, 120, 240\}$ and window length $\in\{240, 600, 1200\}$.
In all experiments, we set $\ell_{reg}$ and $\ell_{sup}$ to be the L1 loss, following the training objective of each phonetic setting, as defined in~\ref{phonetic settings}.

\noindent {\bf Model Configurations.} We evaluate several model configurations with different $U$ and $S$ values. In all experiments we used $kernel=8, depth=5$ and $GLU$ as an activation function.
The pre-trained \ac{ASR} model was taken from the SpeechBrain~\cite{ravanelli2021speechbrain} repository~\footnote{We use the \texttt{asr-crdnn-rnnlm-librispeech} model}; HuBERT was taken from the official implementation under the Fairseq~\cite{ott2019fairseq} repository.

\noindent {\bf Implementation Details.} All models were trained using 4.5 seconds segments with a 0.5 seconds stride. We use a batch size of 16, Adam optimizer with  $\beta_1 = 0.9$ and $\beta_2=0.999$, a learning rate of $3e-4$, and train the models for 300 epochs. Under the \emph{conditioning} setting we concatenate the phonetic features with the encoder output followed by a linear projection layer, and feed it into the LSTM module.

\subsection{Data}
\label{subsec:data}
We train and evaluate our models using the Valentini benchmark~\cite{valentini2017noisy} consisting of 14 male and 14 female speakers with the same accent region and around 400 sentences per speaker, downsampled from 48kHz to 16kHz. We use the standard train-test split, and use speakers p286 and p287 for the validation set. We further apply data augmentation as in~\cite{defossez2020real}, i.e., we apply a random shift between 0 to 0.5 seconds, shuffle the noises within one batch to form new noisy mixtures and apply a band-stop filter, sampled to remove 20\% of the frequencies uniformly in the mel scale.

\subsection{Model Evaluation}
\label{subsec:eval}
We evaluate the model's performance using the following objective measures: i) PESQ: Perceptual evaluation of speech quality, using the wide-band version recommended in ITU-T P.862.2 \cite{recommendation2001perceptual} (from -0.5 to 4.5). ii) CSIG: Mean opinion score (MOS) prediction of the signal distortion attending only to the speech signal \cite{hu2007evaluation} (from 1 to 5).
iii) CBAK: MOS prediction of the intrusiveness of background noise \cite{hu2007evaluation} (from 1 to 5). iv) COVL: MOS prediction of the overall effect \cite{hu2007evaluation} (from 1 to 5).
v) ViSQOL: a signal-based, full-reference, intrusive metric that models human speech quality perception using a spectro-temporal measure of similarity between a reference and a test speech signal \cite{hines2015visqol} (from 1 to 5). We additionally evaluated the Short-Time Objective Intelligibility (STOI) \cite{taal2011algorithm} metric. However, as we observed only negligible changes in results, this metric was omitted from the result tables.

\vspace{-0.1cm}
\subsection{Results}
\label{subsec:res}
The results are summarized in Table~\ref{tab:main} suggest that throughout the experimented configurations, injecting phonetic information consistently improves model performance in most settings under the non-causal setup.
Interestingly, under the causal setup, none of the phonetic settings significantly changed performance in comparison to the baseline model.  
It appears that HuBERT-based phonetic features perform better than ASR-based phonetic features on most settings, where conditioning shows consistent notable margin between the two, for instance 3.72 COVL for HuBERT vs. 3.64 COVL for ASR, under the $H{=}48$ setting.
In general, even considering a simple conditioning linear module, the conditioning setting seems to contribute to model performance the most among the tested phonetic settings.
We hypothesize that due to learning general purpose speech representation, using a masked prediction loss function without any textual supervision, HuBERT captures additional information to phonetic content, e.g., acoustic information, thus outperforming the \ac{ASR} representation.

\subsection{Layer Selection}
\label{subsec:analysis}
We further analyze the impact on performance due to using different conditioning feature vectors drawn from different hidden layers of HuBERT. Specifically, we evaluate: i) each layer independently; ii) an average of all layers; and iii) a learned weighted average of all layers, using Demucs (H=48,U=4,S=4, non-causal) as our speech enhancement model.
Table~\ref{tab:layers} presents the measured metrics with respect to the suggested experiments over layer selection.
The results suggest that injecting phonetic features benefits the model in each of the tested configurations, with one exception of layer 0.
It also follows that the speech enhancement model benefits most when considering a weighted average of HuBERT's hidden layers for conditioning, indicating that relevant information for speech reconstruction is scattered across multiple hidden layers of the conditioning module. 

In addition, the HuBERT paper~\cite{hsu2021hubert} contains an analysis in which the authors evaluate the use of each individual layer with respect to phone normalized mutual information (PNMI), phone purity, and k-means cluster purity metrics. 
Each of which showed higher measures considering layers 5-8 compared to other layers.
As this observation mostly correlates with the results in Table~\ref{tab:layers}, the weight assigned to each layer of HuBERT, visualized in Figure~\ref{fig:pie} , draws a different weight distribution. 
It shows an assignment of higher weight values to lower layers of the model (excluding L0), and not to the mid-high ones, who show higher PNMI values, as one would intuitively expect.
This observation not only support previous indication stating that relevant information is being scattered across different layers, it also highlights the importance of the information held in lower layers which has lower PNMI values, and may imply that shared information is taken into account in the weighted average.

%% file: 04.exp.x-factor.table.tex
\begin{table}[t!]
\centering
\caption{Comparison of HuBERT's hidden layer selection impact over phonetic conditioning results; The setting correlates to the Demucs (H=48,U=4,S=4,non-causal) baseline.}
\vspace{-1em}
\resizebox{0.90\columnwidth}{!}{\begin{tabular}{@{\hskip5pt}l|ccccccc} 
    \toprule
    \bf Layer \# & \bf CBAK & \bf COVL & \bf CSIG & \bf PESQ & \bf VISQOL \\ 
    \midrule
    Baseline & 3.50 & 3.65 & 4.33 & 2.93 & 3.16  \\
    0 & 3.49 & 3.65 & 4.33 & 2.93 &  3.23 \\
    1 & 3.50 & 3.68 & 4.37 & 2.95 &  3.25 \\
    2 & 3.55 & 3.73 & 4.39 & 3.02 &  3.27 \\
    3 & 3.51 & 3.71 & 4.38 & 2.99 &  3.28 \\
    4 & 3.55 & 3.71 & 4.37 & 3.00 &  3.27 \\
    5 & 3.53 & 3.71 & 4.38 & 3.00 &  3.27 \\
    6 & 3.53 & 3.71 & 4.38 & 2.99 &  3.25 \\
    7 & 3.52 & 3.67 & 4.35 & 2.95 &  3.25 \\
    8 & 3.51 & 3.68 & 4.36 & 2.96 &  3.27 \\
    9 & 3.51 & 3.65 & 4.33 & 2.94 &  3.25 \\
    10 & 3.52 & 3.69 & 4.35 & 2.98 & 3.22 \\
    11 & 3.50 & 3.68 & 4.36 & 2.96 & 3.27 \\
    12 & 3.49 & 3.67 & 4.35 & 2.95 & 3.27 \\
    Avg(0-12) & \textbf{3.56} & \textbf{3.76} & \textbf{4.42} & \textbf{3.06} & \bf{3.29}  \\
    Lrn-W-Avg(0-12) & \textbf{3.59} & \textbf{3.78} & \textbf{4.43} & \textbf{3.07} & \bf{3.30} \\
    \bottomrule
    \end{tabular}}
\vspace{-2em}
\label{tab:layers}
\end{table}

%% file: 02.related.tex
\vspace{-0.2cm}
\section{Related Work}
\label{sec:related}

\noindent {\bf Speech Enhancement.}
Traditionally, speech enhancement methods generate either an enhanced version of the magnitude spectrum or produce an estimate of the ideal binary mask (IBM) that is then used to enhance the magnitude spectrum~\cite{ephraim1984speech,hu2006subjective}.
With the rise of deep learning methods there has been a growing interest toward deep neural network based methods for speech enhancement, e.g. \cite{wang2015deep, weninger2015speech}, 
leading to various studies in recent years, such as \cite{phan2020improving, rethage2018wavenet, defossez2020real,fu2021metricgan+}.

\noindent {\bf Self-Supervised Learning.}
\ac{SSL} introduced a methodology by which we obtain supervisory signals from the data itself, thereby enabling the training of large models over huge unlabeled datasets and leading to significant breakthroughs in recent years, e.g. BERT~\cite{devlin2018bert}.
In \cite{schneider2019wav2vec, baevski2019vq} the authors have trained a speech representation model to predict future samples from a given audio signal. HuBERT~\cite{hsu2021hubert} also focuses on speech representation and uses masked prediction similar to \cite{devlin2018bert}. \ac{SSL} models have shown great success in the plenty of discriminate speech tasks ~\cite{yang2021superb}.

\noindent {\bf Phonetic Aware Speech Enhancement.} Models that incorporate phonetic information can utilize either the inclusion of a perceptual loss or knowledge distillation. In~\cite{hsieh2020improving} the authors use a perceptual loss supervision based on wav2vec~\cite{schneider2019wav2vec} encoding, incorporated with the Wasserstein distance metric. Similarly, in~\cite{kim2018unpaired} the authors purposed a pretrained acoustic model to provide supervision during model training. Similar trends were observed in speaker source separation models~\cite{kanda2019guided, von2020multi}. Unlike the above methods, the authors of~\cite{du2020pan} proposed injecting supervised phonetic information by either augmenting the training objective with phonemes prediction or by conditioning a latent representation of the enhancement model. 

%% file: 05.conclusion.tex
\begin{figure}[t!]
\centering
\includegraphics[width=0.95\columnwidth]{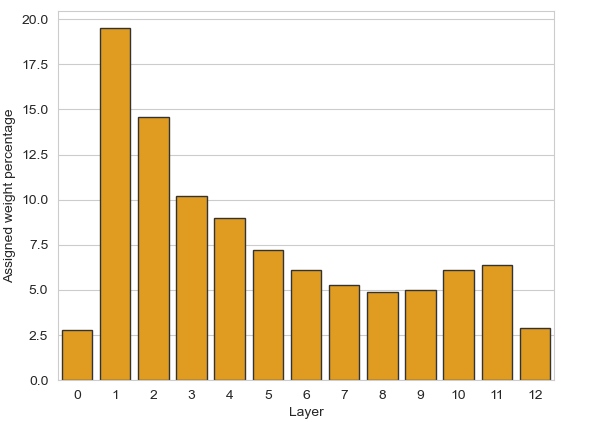}
\vspace{-0.3cm}
\caption{Learned weight distribution of HuBERT's hidden layers coefficients used for phonetic weighted average conditioning; The setting correlates to the Demucs (H=48,U=4,S=4,non-causal) baseline. Similar trends were inspected in all other baselines}
\label{fig:pie}
\vspace{-2em}
\end{figure}

\vspace{-0.2cm}
\section{Conclusions \& Discussion}
\label{sec:conclude}
In this study, we perform a systematic comparison of three different settings for incorporating phonetic features, \emph{regularization}, \emph{supervision} and \emph{conditioning}, considering causal and non-causal setups with respect to the speech enhancement task.

The use of phonetic features shows an improvement in model performance for speech enhancement in the non-causal setup. Specifically, the learned weighted average conditioning setting shows significant improvement, compared to other phonetic settings that are far more common in the speech enhancement domain.
This study presents several observations worth exploring which we intend to address in future work. 

First, a simple single linear layer conditioning of a learned weighted average shows an average improvement of $3.5\%$ over the observed metrics. This result leaves room for further exploring the conditioning of module architectures, thereby expanding the impact of the conditioning setting. Second, the effect of incorporating phonetic features in the causal setup is surprisingly negligible. Intuitively, we would expect phonetic features to improve performance, which leaves room for exploration and analysis of this phenomena. Last, the choice of a phonetic model determines the feature representation. Our setup allows isolated observations, and could therefore serve as a platform for in-depth analysis of various speech representations drawn from different models.